\documentclass[aps,pre,twocolumn,floats,floatfix,english]{revtex4}
\usepackage[T1]{fontenc}
\usepackage[latin1]{inputenc}
\usepackage{amsmath}
\usepackage{setspace}
\usepackage{amssymb}

\makeatletter

\providecommand{\LyX}{L\kern-.1667em\lower.25em\hbox{Y}\kern-.125emX\@}

\usepackage{setspace}
\usepackage{epsfig}
\usepackage{babel}
\makeatother

\begin{document}
\title{Dynamics of market correlations: Taxonomy
and portfolio analysis}

\author{J.-P. Onnela, A. Chakraborti, K. Kaski}
\affiliation{Laboratory of Computational Engineering, 
Helsinki University of Technology, P.O. Box 9203, FIN-02015 HUT, Finland}
\author{J. Kertész}
\affiliation{Department of Theoretical Physics, Budapest University 
of Technology \& Economics, Budafoki út 8, H-1111, Budapest, Hungary}
\affiliation{Laboratory of Computational Engineering, 
Helsinki University of Technology, P.O. Box 9203, FIN-02015 HUT, Finland} 
\author{A. Kanto}
\affiliation{Department of Quantitative Methods in Economics and 
Management Science, Helsinki School of Economics, P.O.Box 1210, 
FIN-00101 Helsinki, Finland}

\begin{abstract}
The time dependence of the recently introduced minimum spanning 
tree description of correlations between stocks, called the 
``asset tree'' have been studied to reflect the economic 
taxonomy. The nodes of the tree are identified with stocks 
and the distance between them is a unique function of the 
corresponding element of the correlation matrix. 
By using the concept of a central vertex, chosen as the most 
strongly connected node of the tree, an important characteristic 
is defined by the mean occupation layer (MOL). During crashes 
the strong global correlation in the market manifests itself 
by a low value of MOL. The tree seems to have a scale free 
structure where the scaling exponent of the degree distribution 
is different for `business as usual' and `crash' periods. 
The basic structure of the tree topology is very robust with 
respect to time. We also point out that the diversification 
aspect of portfolio optimization results in the fact that 
the assets of the classic Markowitz portfolio are always 
located on the outer leaves of the tree. Technical aspects 
like the window size dependence of the investigated quantities 
are also discussed.
\end{abstract}

\maketitle 
\section{Introduction}

In spite of the traditional wisdom ``Money does not grow on trees'', 
here we wish to show that the concept of trees (graphs) have potential 
applications in financial market
analysis. This concept was recently introduced by Mantegna as a method 
for finding a hierarchical arrangement of stocks through studying 
the clustering of companies by using correlations of asset returns \cite{Man1}.
With an appropriate metric, based on the correlation matrix, a fully
connected graph was defined in which the nodes are companies, or 
stocks, and the `distances' between them are obtained from the 
corresponding correlation coefficients. The minimum spanning tree (MST) 
was generated from the graph by selecting the most important 
correlations and it is used to identify clusters of companies. 

In this paper, we study the time dependent properties of the minimum 
spanning tree and call it a `dynamic asset tree'.
It should be mentioned that several attempts
have been made to obtain clustering from the huge correlation matrix,
like the Potts super paramagnetic method \cite{Kull}, a method based 
on the maximum likelihood \cite{Mar} or the comparison of the
eigenvalues with those given by the random matrix theory \cite{Lal}. 
We have chosen the MST because of its uniqueness and simplicity. 
The different methods are compared in \cite{Mar}.

Financial markets are often characterized as evolving complex 
systems \cite{santafe}. The evolution is a reflection of the 
changing power structure in the market and it manifests the passing of 
different products and product generations, new technologies, 
management teams, alliances and partnerships, among many other factors. 
This is why exploring the asset tree \emph{dynamics} can provide us 
new insights to the market. We believe that dynamic asset trees can 
be used to simplify this complexity in order to grasp the essence 
of the market without drowning in the abundance of information. 
We aim to derive intuitively understandable measures, which can be 
used to characterize the market taxonomy and its state. A further
characterization of the asset tree is obtained by studying its degree
distribution \cite{Van}.
We will also study the robustness of tree topology and the consequences of 
the market events on its structure. The minimum spanning tree, as 
a strongly pruned representative of asset correlations, is found to 
be robust and descriptive of stock market events.

Furthermore, we aim to apply dynamic asset trees in the field 
of portfolio optimization. Many attempts have been made to solve 
this central problem from the classical approach of 
Markowitz \cite{Mark} to more sophisticated treatments, including 
spin glass type studies \cite{Gall}. In all the attempts to solve 
this problem, correlations between asset prices play a crucial 
role and one might, therefore, expect a connection between dynamic 
asset trees and the Markowitz portfolio optimization scheme. We 
demonstrate that although the topological structure of the tree 
changes with time, the companies of the minimum risk Markowitz portfolio 
are always located on the outer leaves of the tree. Consequently, 
asset trees in addition to their ability to form economically meaningful 
clusters, could potentially contribute to the portfolio optimization 
problem. Then with a lighter key one could perhaps say that 
``some money may grow on trees'', after all.

The paper is organized as follows. In Section 2  
we introduce the data, discuss some properties of asset return 
correlation distributions and construct and characterize trees.
Section 3 deals with tree occupation and central vertex considerations, 
followed with Section 4 which addresses the important question 
of economic meaningfulness of tree clusters. Then Section 5 is devoted 
to the study of the scale free character of the asset trees. Section 6 
deals with tree evolution through the concepts of two different 
types of survival ratios, which can be used to describe decaying 
of connections and determine tree half-lives. In the subsequent 
Section 7, we investigate how asset trees can contribute to the 
portfolio optimization problem. Finally, in Section 8, we draw 
conclusions and summarize our findings.

\section{Return correlations and dynamic asset trees}

The financial market, for the largest part in this paper, refers to 
a set of data commercially available from the Center for Research 
in Security Prices (CRSP) of the University of Chicago Graduate School 
of Business. Here We will study the split-adjusted daily closure prices 
for a total of $N=477$ stocks traded at the New York Stock Exchange (NYSE) 
over the period of 20 years, from 02-Jan-1980 to 31-Dec-1999. This 
amounts a total of 5056 price quotes per stock, indexed by time variable 
$\tau = 1, 2, \ldots, 5056$. For analysis and smoothing purposes, the 
data is divided time-wise into $M$ \emph{windows} $t=1,\, 2,...,\, M$ 
of width $T$ corresponding to the number of daily returns included in 
the window. Several consecutive windows overlap with each other, the 
extent of which is dictated by the window step length parameter 
$\delta T$, describing the displacement of the window, measured also 
in trading days. The choice of window width is a trade-off between too 
noisy and too smoothed data for small and large window widths, 
respectively. The results presented in this paper were calculated from 
monthly stepped four-year windows, i.e. $\delta T \approx 20.8$ days and 
$T=1000$ days. We have explored a large scale of different values for both 
parameters, and the given values were found optimal \cite{jpo}. With 
these choices, the overall number of windows is $M=195$.

In order to investigate correlations between stocks we first denote 
the closure price of stock $i$ at time $\tau$ by $P_{i}(\tau)$ 
(Note that $\tau$ refers to a date, not a time window). We focus 
our attention to the logarithmic return of stock $i$, given by 
$r_{i}(\tau)=\ln P_{i}(\tau)-\ln P_{i}(\tau-1)$ which, for a sequence 
of consecutive trading days, i.e. those encompassing the given window 
$t$, form the return vector $\boldsymbol r_{i}^t$. In order to 
characterize the synchronous time evolution of assets, we use the equal 
time correlation coefficients between assets $i$ and $j$ defined as

\begin{equation}
\rho _{ij}^t=\frac{\langle \boldsymbol r_{i}^t \boldsymbol r_{j}^t \rangle -\langle \boldsymbol r_{i}^t \rangle \langle \boldsymbol r_{j}^t \rangle }{\sqrt{[\langle {\boldsymbol r_{i}^t}^{2} \rangle -\langle \boldsymbol r_{i}^t\rangle ^{2}][\langle {\boldsymbol r_{j}^t}^{2} \rangle -\langle \boldsymbol r_{j}^t \rangle ^{2}]}},
\end{equation}

\noindent where $\left\langle ...\right\rangle $ indicates a time 
average over the consecutive trading days included in the return 
vectors. Due to Cauchy-Schwarz inequality, these correlation 
coefficients fulfill the condition $-1\leq \rho _{ij}\leq 1$ 
and form an $N\times N$ correlation matrix $\mathbf{C}^t$, which 
serves as the basis of dynamic asset trees to be discussed later.

\begin{figure}
\epsfig{file=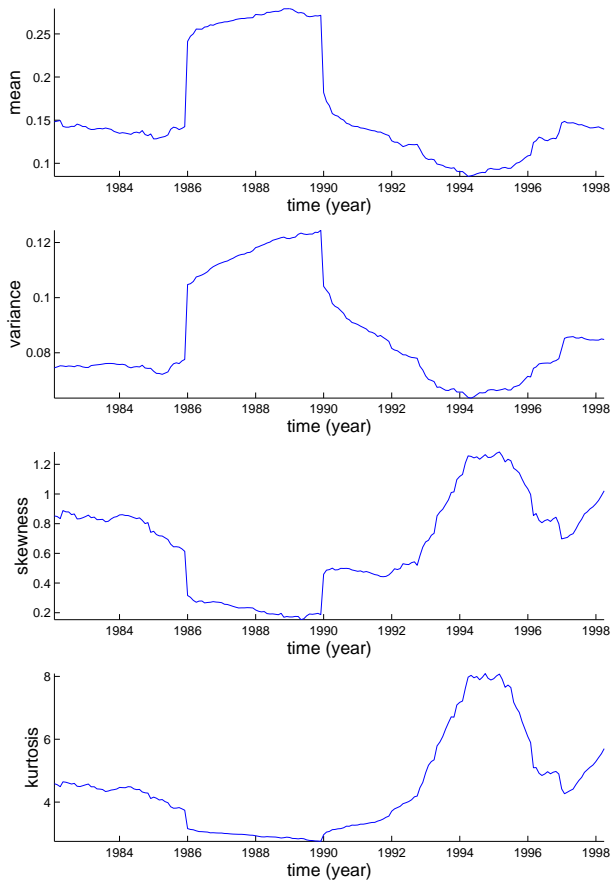,width=3.2in }
\caption{
The mean, variance, skewness and kurtosis of the correlation coefficients as functions of time.}
\label{fig:mvsk}
\end{figure}

Let us first characterize the correlation coefficient distribution 
by its first four moments and their correlations with one another. 
The first moment is the \emph{mean correlation coefficient} defined as 

\begin{equation}
\bar{\rho}(t)=\frac{1}{N(N-1)/2}\sum _{\rho_{ij}^t \in \mathbf{C}^t}\rho _{ij}^t, 
\end{equation}

\noindent where we consider only the non-diagonal $(i \neq j)$ elements $\rho _{ij}^t$ of the upper (or lower) triangular matrix. 
We also evaluate the higher order moments 
for the correlation coefficients, so that the variance is 
\begin{equation}
\lambda _{2}(t)=\frac{1}{N(N-1)/2}\sum_{(i,j)} (\rho _{ij}^t-\bar{\rho }^t)^{2},
\end{equation} 
the skewness is 
\begin{equation}
\lambda _{3}(t)=\frac{1}{N(N-1)/2}\sum_{(i,j)} (\rho _{ij}^t-\bar{\rho }^t)^{3}/\lambda _{2}^{3/2}(t), 
\end{equation}
and the kurtosis is 
\begin{equation}
\lambda _{4}(t)=\frac{1}{N(N-1)/2}\sum_{(i,j)} (\rho _{ij}^t-\bar{\rho }^t)^{4}/\lambda _{2}^{2}(t).
\end{equation} 
The mean, variance, skewness and kurtosis of the correlation coefficients 
are plotted as functions of time in Figure \ref{fig:mvsk}. 

In this figure the effect 
and repercussions of Black Monday (October 19, 1987) are clearly 
visible in the behavior of all these quantities. For example, the mean 
correlation coefficient is clearly higher than average on the interval 
between 1986 and 1990. The length of this interval corresponds to the 
window width $T$, and Black Monday coincides with the mid-point of the 
interval \cite{bali}. The increased value of the mean correlation 
is in accordance with the observation by Drozdz et al. \cite{Dro}, who 
found that the maximum eigenvalue of the correlation matrix, which 
carries most of the correlations, is very large during 
market crashes. We also investigated whether these four 
different measures are correlated, as seems clear from the figure. 
For this we determined the Pearson's linear and Spearman's rank-order 
correlation coefficients, which between the mean and variance turned out 
to be 0.97 and 0.90, and between skewness and kurtosis 0.93 and 0.96, 
respectively. Thus the first two and the last two measures are very 
strongly correlated.

We now move on to construct an asset tree. For this we use the 
non-linear transformation $d_{ij}=\sqrt{2(1-\rho _{ij})}$ to 
obtain distances with the property $2\geq d_{ij}\geq 0$, forming an 
$N\times N$ distance matrix $\mathbf{D}^t$. At this point an 
additional hypothesis about the topology of the metric space 
is required. The working hypothesis is that a useful space for 
linking the stocks is an \emph{ultrametric space}, i.e., a 
space where all distances are ultrametric. This hypothesis is 
motivated \emph{a posteriori} by the finding that the associated 
taxonomy is meaningful from an economic point of view. The concept 
of ultrametricity is discussed in detail by Mantegna \cite{Man1}, 
while the economic meaningfulness of the emerging taxonomy is 
addressed later in this paper. Out of the several possible 
ultrametric spaces, the subdominant ultrametric is opted for 
due to its simplicity and remarkable properties. In practice, 
it is obtained by using the distance matrix $\mathbf{D}^t$ 
to determine the minimum spanning tree (MST) of the distances, 
according to the methodology of \cite{Man1}, denoted $\mathbf{T}^t$. 
This is a simply connected graph that connects all $N$ nodes of 
the graph with $N-1$ edges such that the sum of all edge weights, 
$\sum _{d_{ij}^t \in \mathbf{T}^t}d_{ij}^t$, is minimum. 
(Here time (window) dependence of the tree is emphasized 
by the addition of the superscript $t$ to the notation.) 
Asset trees constructed for different time windows are not 
independent from each other, but form a series through time. 
Consequently, this multitude of trees is interpreted as a 
sequence of evolutionary steps of a single \emph{dynamic asset tree}.

\begin{figure}
\epsfig{file=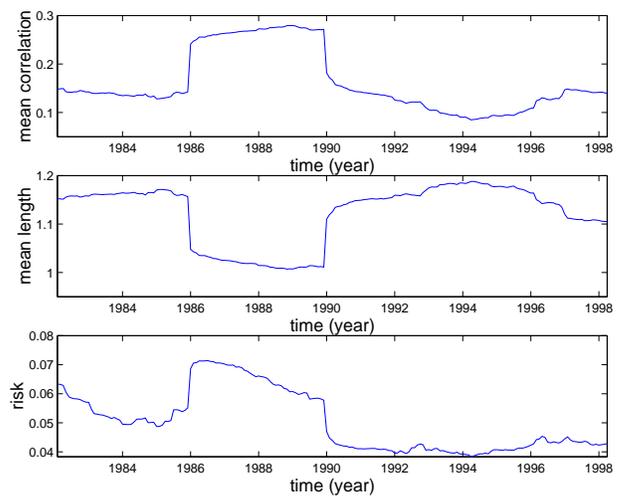,width=3.2in }
\caption{
Plots of (a) the mean correlation coefficient $\bar{\rho}(t)$, 
(b) the normalized tree length $L(t)$ and (c) the risk of the 
minimum risk portfolio, as functions of time.
}
\label{clr}
\end{figure}

As a simple measure of the temporal state of the market 
(the asset tree) we define the \emph{normalized tree length} as

\begin{equation}
L(t)=\frac{1}{N-1}\sum _{d_{ij}^t \in \mathbf{T}^t}d_{ij}^t,\end{equation}

\noindent where $t$ again denotes the time at which the tree is 
constructed, and $N-1$ is the number of edges present in the MST. 
The normalized tree length is depicted in Figure \ref{clr}. 

As expected and as the plots show, the mean correlation coefficient
and the normalized tree length are very strongly anti-correlated. 
Pearson's linear correlation between 
the mean correlation coefficient $\bar{\rho}(t)$ and normalized tree 
length $L(t)$ is -0.98, and Spearman's rank-order 
correlation coefficient is -0.92, thus both indicating very 
strong anti-correlation. Anti-correlation is to be expected in 
view of how the distances $d_{ij}$ are constructed from correlation 
coefficients $\rho_{ij}$. However, the extent of this anti-correlation 
is different for different input variables and is lower if, say, 
daily transaction volumes are studied instead of daily closure 
prices \cite{inprep}. 

It should be noted that in constructing the minimum spanning tree, 
we are effectively reducing the information space from $N(N-1)/2$ 
separate correlation coefficients to $N-1$ tree edges, in other words, 
compressing the amount of information dramatically. This follows because 
the correlation matrix $\mathbf{C}^t$ and distance matrix 
$\mathbf{D}^t$ are both $N\times N$ dimensional, but due to 
their symmetry, both have $N(N-1)/2$ distinct upper (or lower) 
triangle elements, while the spanning tree has only $N-1$ edges. 
So, in moving from correlation or distance matrix 
to the asset tree, we have pruned the system from $N(N-1)/2$ 
to $N-1$ elements of information. This, of course, raises the key question 
of information theory, whether essential information is lost in the 
reduction. As the above examination of the mean correlation coefficient 
and normalized tree length shows, the fact that the two 
measures are strongly anti-correlated testifies to the success of 
the pruning process. Consequently, one is justified to contemplate 
the minimum spanning tree as a strongly reduced representative of 
the whole correlation matrix, which bears the essential information 
about asset correlations. 

As further evidence that the MST retains the salient features of 
the stock market, it is noted that the 1987 market crash can be 
quite accurately seen in Figure \ref{clr}. The fact that the market 
, during crash, is moving together is thus manifested in two ways. 
First, the ridge in the plot of the mean correlation coefficient 
in Figure \ref{clr}(a) indicates that the whole market is 
exceptionally strongly correlated. Second, the corresponding 
well in the plot of the normalized tree length 
in Figure \ref{clr}(b) shows how this is reflected in considerably 
shorter than average length of the tree so that the tree, on 
average, is very tightly packed. Upon letting the window width 
$T \to 0$, the two sides of the ridge converge to a single date, 
which coincides with Black Monday \cite{bali}.

\section{Tree occupation and central vertex}

Next we focus on characterizing the spread of nodes on the tree. 
In order to do so, we introduce the quantity of \emph{mean 
occupation layer} as 

\begin{equation}
l(t,v_c)=\frac{1}{N}\sum _{i=1}^{N}\mathop {\mathrm{lev}}(v_{i}^{t}),\end{equation}

\noindent where $\mathop {\mathrm{lev}}(v_{i})$ denotes the level 
of vertex $v_{i}$. The levels, not to be confused with the 
distances $d_{ij}$ between nodes, are measured in natural numbers 
in relation to the \emph{central vertex} $v_c$, whose level is 
taken to be zero. Here the mean occupation layer indicates the 
layer on which the mass of the tree, on average, is conceived 
to be located.  
 
\begin{figure}
\epsfig{file=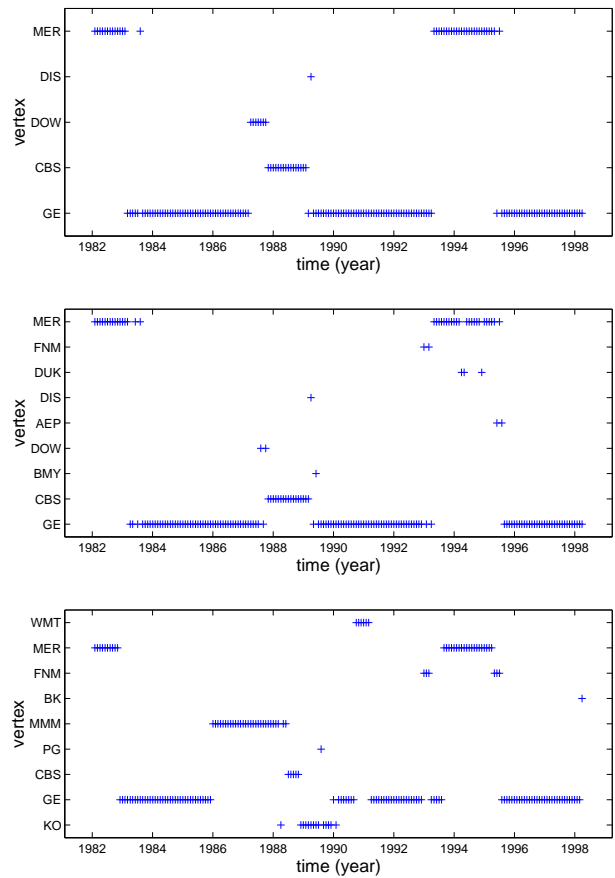,width=3.2in }
\caption{
Central vertices according to (a) vertex degree criterion, 
(b) weighted vertex degree criterion and (c) center of mass criterion.
}
\label{cvs}
\end{figure}

Let us now examine the central vertex in more detail, as the 
understanding of the concept is a prerequisite for interpreting 
mean occupation layer results, to follow shortly. The central 
vertex is considered the parent of all other nodes in the tree, 
also known as the root of the tree. It is used as the reference 
point in the tree, against which the locations of all other nodes 
are relative. Thus all other nodes in the tree are children of 
the central vertex. Although there is arbitrariness in the 
choice of the central vertex, we propose that it is central, 
or important, in the sense that any change in its price strongly 
affects the course of events in the market on the whole. We
propose three alternative definitions have emerged for the central 
vertex in our studies, all yielding similar and, in most cases, 
identical outcomes. The first and second definitions of the 
central vertex are local in nature. The idea here is to find 
the node that is most strongly connected to its nearest neighbors. 
According to the the first definition, this is the node with 
the highest \emph{vertex degree}, i.e. the number of edges 
which are incident with (neighbor of) the vertex. The obtained 
results are shown in Figure \ref{cvs}.

The \emph{vertex degree 
criterion} leads to General Electric (GE) dominating 67.2\% of the 
time, followed by Merrill Lynch (MER) at 20.5\% and CBS at 
8.2\%. The combined share of these three vertices is 95.9\%. 
The second definition, a modification of the first, defines the 
central vertex as the one with the highest sum of those 
correlation coefficients that are associated with the incident 
edges of the vertex. Therefore, whereas the first definition 
weighs each departing node equally, the second gives more weight 
to short edges, since a high value of $\rho_{ij}$ corresponds 
to a low value of $d_{ij}$. This is reasonable, as short connections 
link the vertex more tightly to its neighborhood than long 
ones (the same principle employed in constructing the spanning 
tree). This \emph{weighted vertex degree criterion} results in GE 
dominating 65.6\% of the cases, followed by MER at 20.0\% and 
CBS at 8.7\%, the share of the top three being 94.3\%. The 
third definition deals with the global quantity of 
\emph{center of mass}. In considering a tree $\mathbf{T}^t$ at 
time $t$, the vertex $v_i$ that produces the lowest value for mean 
occupation layer $l(t,v_i)$ is the center of mass, given that 
all nodes are assigned an equal weight and consecutive layers 
(levels) are at equidistance from one another, in accordance 
with the above definition. With this  
\emph{center of mass criterion} we find that the most dominant 
company, again, is GE, as it is 52.8\% of the time the centre 
of mass, followed by MER at 15.4\% and Minnesota Mining 
\& MFG at 14.9\%. These top three candidates constitute 83.1\% 
of the total. Should the weight of the node be made proportional 
to the size (e.g. revenue, profit etc.) of the company, it is 
obvious that GE's dominance would increase.

As Figure \ref{cvs} shows, the three alternative definitions for 
the central vertex lead to very similar results. The vertex degree 
and the weighted vertex degree criteria coincide 91.8\% of the time. 
In addition, the former coincides with center of mass 66.7\% and 
the latter 64.6\% of the time, respectively. Overall, the three 
criteria yield the same central vertex in 63.6\% of the cases, 
indicating considerable mutual agreement. The existence of a 
meaningful center in the tree is not a trivial issue, and neither 
is its coincidence with the center of mass. However, since the criteria 
applied, present a mixture of both local and global approaches, 
and the fact that they coincide almost 2/3 of the time, does indicate 
the existence of a well-defined center in the tree. The reason 
for the coincidence of the criteria seems clear, intuitively 
speaking. A vertex with a high vertex degree, the central vertex 
in particular, carries a lot of weight around it 
(the neighboring nodes), which in turn may be highly connected 
to others (to their children) and so on. 
Two different interpretations may be given to these results. 
One may have either (i) static (fixed at all times) or (ii) 
dynamic (updated at each time step) central vertex. If the 
first approach is opted for, the above evidence well 
substantiates the use of GE as the central vertex. In the 
second approach, the results will vary somewhat depending 
on which of the three criteria is used in determining the 
central vertex.

\begin{figure}
\epsfig{file=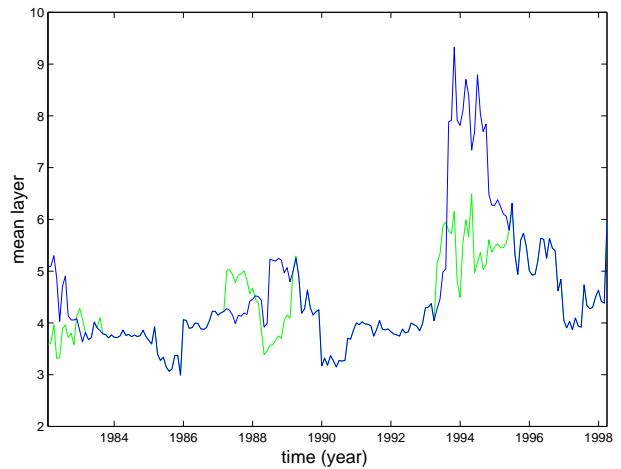,width=3.2in }
\caption{
Plot of mean occupation layer $l(t,v_{c})$ as a function of time, 
with static and dynamic central vertices.
}
\label{mol}
\end{figure}

The mean occupation layer $l(t)$ is depicted in Figure \ref{mol}, 
where also the effect of different central vertices is demonstrated. 
The blue curve results from the static central vertex, i.e. GE, and 
the green one to dynamic central vertex evaluated using the vertex 
degree criterion. The two curves coincide where only the blue curve 
is drawn. This is true most of the time, as the above central vertex 
considerations lead us to expect. The two dips at 1986 and 1990, 
located symmetrically at half a window width from Black Monday, 
correspond to the topological shrinking of the tree associated 
with the famous market crash of 1987 \cite{bali}. Roughly between 
1993 and 1997 $l(t)$ reaches very high values, which is in concordance 
with our earlier results obtained for a different set of data 
\cite{short}. High values of $l(t)$ are considered to reflect 
a finer market structure, whereas in the other 
extreme low dips are connected to market crashes, where the behavior 
of the system is very homogeneous. The finer structure may result 
from general steady growth in asset prices during that period as can be seen, 
for example, from the S\&P 500 index.

\section{Tree clusters and their economic meaningfulness}

As mentioned earlier, Mantegna's idea of linking stocks in an 
ultrametric space was motivated \emph{a posteriori} by the 
property of such a space to provide a meaningful economic 
taxonomy. We will now explore this issue further, as the 
meaningfulness of the emerging economic taxonomy is the key 
justification for the use of the current methodology. 
In \cite{Man1}, Mantegna examined the meaningfulness of 
the taxonomy by comparing the grouping of stocks in the 
tree with a third party reference grouping of stocks by 
their industry etc. classifications. In this case, the 
reference was provided by Forbes\cite{forbes}, which 
uses its own classification system, assigning each stock with 
a sector (higher level) and industry (lower level) category.. 

\begin{figure}
\epsfig{file=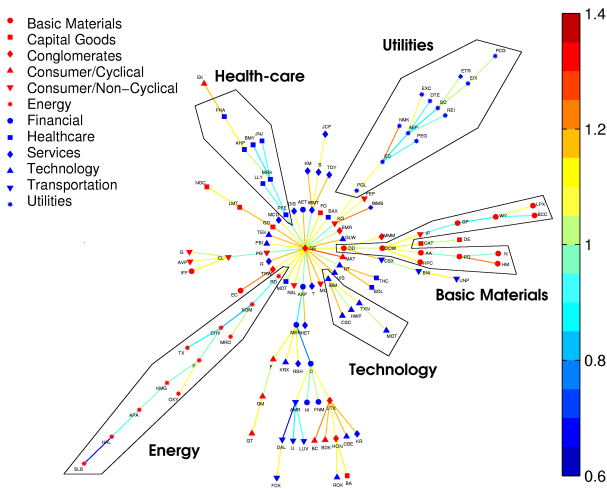,width=3.2in }
\caption{
Snapshot of a dynamic asset tree connecting the examined 116 stocks 
of the S\&P 500 index. The tree was produced using four-year 
window width and it is centered on January 1, 1998. Business 
sectors are indicated according to Forbes, \emph{http://www.forbes.com}. 
In this tree, General Electric (GE) was used as the central 
vertex and eight layers can be identified.
}
\label{samplegraph}
\end{figure}

In order to visualize the grouping of stocks, we constructed a 
sample asset tree for a smaller dataset, shown in Figure \ref{samplegraph}. 
This was obtained by studying our previous dataset \cite{short}, 
which consists of 116 S\&P 500 stocks, extending from the beginning 
of 1982 to the end of 2000, resulting in a total of 4787 price 
quotes per stock \cite{supp}. The window width was set at $T=1000$, and the 
shown sample tree is located time-wise at $t=t^*$, corresponding 
to 1.1.1998. The stocks in this dataset fall into 12 \emph{sectors}, 
which are Basic Materials, Capital Goods, Conglomerates, 
Consumer/Cyclical, Consumer/Non-Cyclical, Energy, Financial, 
Healthcare, Services, Technology, Transportation and Utilities. 
The sectors are indicated in the tree with different markers, 
while the industry classifications are omitted for reasons of clarity.

Before evaluating the economic meaningfulness of grouping stocks, 
we wish to establish some terminology. We use the term sector 
exclusively to refer to the given third party classification 
system of stocks. The term \emph{branch} refers to a subset of 
the tree, to all the nodes that share the specified common parent. 
In addition to the parent, we need to have a reference point 
to indicate the generational direction (i.e. who is who's parent) 
in order for a branch to be well defined. Without this reference 
there is absolutely no way to determine where one branch ends 
and the other begins. In our case, the reference is the central 
node. There are some branches in the tree, in which most of the 
stocks belong to just one sector, indicating that the branch is 
fairly homogeneous with respect to business sectors. This finding 
is in accordance with those of Mantegna \cite{Man1}, although 
there are branches that are fairly heterogeneous, such as the 
one extending directly downwards from the central vertex, see 
Figure \ref{samplegraph}. 

Since the grouping of stocks is not perfect at the branch level, 
we define a smaller subset whose members are more homogeneous as 
measured by the uniformity of their sector classifications. The 
term \emph{cluster} is defined, broadly speaking, as a subset 
of a branch, but a more accurate definition is based on the 
following four rules. (i) A cluster is named after the cluster 
parent, which is the node in the cluster closest to the central 
vertex and it is the starting node of the cluster. The cluster is 
named after the business sector of the cluster parent. This is 
why, for example, Utilities cluster starts from PGL and not 
from KO. (ii) If there are more than one potential cluster parent, 
the one resulting in the most complete cluster is chosen 
as the cluster parent. The nodes that are left outside the 
formed cluster are considered outliers. (iii) Only those edges 
that are required to connect the cluster are included. Therefore, 
for example, in the Basic Materials cluster, the edges DOW-IP 
and IP-GP are counted, even though IP is not a Basic Materials 
company, but it is needed to render the cluster connected. 
(iv) If there are nodes in a cluster which do not belong there, 
and they do not have children that belong to the cluster either, 
they are not included. For example, again in the Basic Materials 
cluster, edges DD-CSX-BNI-UNP are not  counted as they do not 
have children that belong to the Basic Materials  sector, 
although the parent DD is a member of the cluster. 
Consequently, CSX, BNI and UNP are not included in the 
Basic Materials cluster. 

Let us now examine some of the clusters that have been formed 
in the sample tree. We use the terms \emph{complete} and \emph{incomplete} 
to describe, in rather strict terms, the success of clustering. 
A complete cluster contains all the companies of the studied 
set belonging to the corresponding business sector, so that none 
are left outside the cluster. In practice, however, clusters are 
mostly incomplete, containing most, but not all, of the companies 
of the given business sector, and the rest are to be found somewhere 
else in the tree. Only the Energy cluster was found complete, but 
many others come very close, typically missing just one or two 
members of the cluster. 

Building upon the normalized tree length concept, we can 
characterize the strength of clusters in a similar manner, as they 
are simply subsets of the tree. These clusters, whether complete 
or incomplete, are characterized by the \emph{normalized cluster length}, 
defined for a cluster $c$ as follows

\begin{equation}
L_{c}(t)=\frac{1}{N_c}\sum_{d_{ij}^t \in c} d_{ij}^t,
\label{eq:mcd}
\end{equation}

\noindent where $N_c$ is the number of stocks in the cluster. 
This can be compared with the normalized tree length, 
which for the sample tree in Figure \ref{samplegraph} at time $t^*$ 
is $L(t^*) \approx 1.05$. A full account of the results is to 
be found in Appendix A, but as a short summary of results we state the 
following. The Energy companies form the most tightly packed 
cluster resulting in $L_{\text{Energy}}(t^*) \approx 0.92$, followed 
by the Health-care cluster with $L_{\text{Health-care}}(t^*) \approx 0.98$. 
For the Utilities cluster we 
have $L_{\text{Utilities}}(t^*) \approx 1.01$ and for the diverse Basic 
Materials cluster $L_{\text{Basic materials}}(t^*) \approx 1.03$. 
Even though the Technology cluster has the fewest number of members, 
its mean distance is the highest of the examined groups of clusters 
being $L_{\text{Technology}}(t^*) \approx 1.07$. Thus, most clusters 
seem to be more tightly packed than the tree on average.

One could find and examine several other clusters in the tree, 
but the ones that were identified are quite convincing. The minimum 
spanning tree, indeed, seems to provide a taxonomy that is well 
compatible with the sector classification provided by an outside 
institution, Forbes in this case. This is a strong vote for the 
use of the current methodology in stock market analysis. Some further 
analysis of the identified clusters is presented in Appendix A.

There are, however, some observed deviations to the classification,
which call for an explanation. For them the following points are
raised. 
(i) Uncertainty in asset prices in the minds of investors causes 
some seemingly random price fluctuations to take place, and this 
introduces ``noise'' in the correlation matrix. Therefore, it is 
not reasonable to expect a one-to-one mapping between business 
sectors and MST clusters. 
(ii) Business sector definitions are not unique, but vary by 
the organization issuing them. In this work, we used the 
classification system by Forbes \cite{forbes}, where the studied 
companies are divided into 12 business sectors and 51 industries. 
Forbes has its own classification principle, based on company 
dynamics rather than size alone.
Alternatively, one could have used, say, the Global Industry 
Classification Standard (GICS), released on January 2, 2001, by 
Standard \& Poor's \cite{sp}. Within this framework, companies 
are divided into 10 sectors, 23 industry groups, 59 industries 
and 122 sub-industries. Therefore, the classification system 
clearly makes a difference, and there are discrepancies even at 
the topmost level of business sectors amongst different systems. 
(iii) Historical price time series is, by definition, old. 
Therefore, one should use contemporary definitions for 
business sectors etc., as those most accurately characterize 
the company. Since these were not available to the authors, the 
classification scheme by Forbes was used. The error caused by 
this approach varies for different companies.
(iv) In many classification systems, 
companies engaged in substantially different business 
activities are classified according to where the majority of 
revenues and profits comes from. For highly diversified companies, 
these classifications are more ambiguous and, therefore, less 
informative. As a consequence, classification of these types 
of companies should be viewed with some skepticism. 
This problem has its roots in the desire to categorize companies 
by a single label, and the approach fails where this division is unnatural.
(v) Some cluster outliers can be explained through the MST clustering 
mechanism, which is based on correlations between asset returns. 
Therefore, one would expect, for example, investment banks to be 
grouped with their investments rather than with other similar 
institutions. Through portfolio diversification, these banks 
distance themselves from the price fluctuations (risks) of a 
single business sector. Consequently, it would be more surprising 
to find a totally homogeneous financial cluster than a fairly 
heterogeneous one currently observed. 
(vi) The risks imposed on 
the companies by the external environment vary in their degree of 
uniformity from one business sector to another. For example, 
companies in the Energy sector (price of their stocks) are prone 
to fluctuations in the world market price of oil, whereas it is 
difficult to think of one factor having equal influence on, say, 
companies in the Consumer/Non-cyclical business sector. This 
uniformity of external risks influences the stock price of these 
companies, in coarse terms, leading to their more complete 
clustering than that of companies facing less uniform external risks.
In conclusion, regarding all the above listed factors, the success 
of the applied method in identifying market taxonomy is remarkable.

\section{Scale Free Structure of the Asset Tree}

So far we have characterized the asset tree as an important subgraph
of the fully connected graph derived from all the elements of the
connectivity matrix. Since the asset tree is expected to reflect some
aspects of the market and its state, it is therefore of interest to
learn more about its structure.  During the last few years, much
attention has been devoted to the degree distribution of graphs. It
has become clear that the so called scale free graphs, where this
distribution obeys a power law, are very frequent in many fields,
ranging from human relationships through cell metabolism to the
Internet \cite{Bar,Dor}. Scale free trees have also been extensively
studied (see e.g., \cite{Sza}).  Recently, examples for scale free
networks in economy and finance have been found \cite{Van,Mar2,Bar2}.

Vandewalle et al. \cite{Van} found scale free behavior
for the asset tree in a limited (one year, 1999) time window for
6358 stocks traded at the NYSE, NASDAQ and AMEX. They proposed the
distribution of the vertex degrees $f(n)$ to follow a power law
behavior:
\begin{equation}
f(n) \sim n ^{-\alpha},
\end{equation}
with the exponent $\alpha \approx 2.2$. This exponent implies that 
the second moment of the distribution would diverge in the infinite 
market limit, or in other words, the second moment of the distribution 
is always dominated by the rare but extremely highly connected
vertices.

\begin{figure}
\epsfig{file=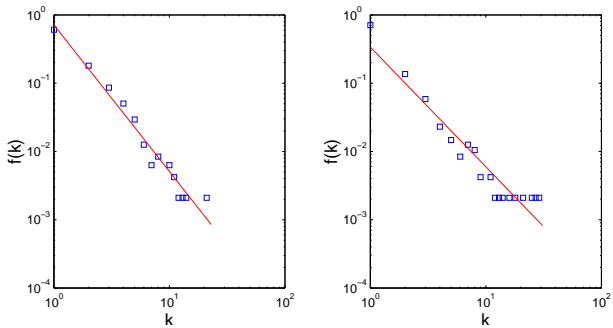,width=3.2in }
\caption{
Typical plots of vertex degree for normal (left) and crash 
topology (right), for which the exponents and goodness of fit 
are $\alpha \approx -2.15$, $R^2 \approx 0.96$ and 
$\alpha \approx -1.75$, $R^2 \approx 0.92$, respectively. 
The plot on the left is centered at 28.2.1994 and the right 
one at 1.5.1989, and for both $T=1000$.
}
\label{vertex_degrees}
\end{figure}

Our aim here is to study the property of scale freeness in the light 
of asset tree dynamics. First, we conclude that the asset tree has, 
most of the time, scale free properties with a rather robust exponent 
$\alpha \approx -2.1 \pm 0.1$ for normal topology (i.e. outside crash 
periods of 'business as usual'), a result close to that given 
in \cite{Van}. For most of the time the distribution behaves in a 
universal manner, meaning that the exponent $\alpha$ is a constant 
within the error limits. However, when the behavior of the market 
is not 'business as usual' (i.e. within crash periods), the exponent 
also changes, although the scale free character of the tree is still 
maintained. For the Black Monday period, we have 
$\alpha \approx -1.8 \pm 0.1$. This result is in full agreement with 
the observation of the shrinking of the tree 
during market crashes, which is accompanied by an increase in the 
degree, thus explaining the higher value of the exponent. The 
observation concerning the change in the value of the exponent 
for normal and crash period is exemplified in Figure \ref{vertex_degrees}.

When fitting the data, in many cases we found one or two outliers, 
i.e. vertices whose degrees did not fit to the overall power law 
behavior since they were much too high. In all cases these stocks 
corresponded either to the highest connected node (i.e. the central 
vertex) or were nodes with very high degrees. This result suggest 
that it could be useful to handle these nodes with special care,  
thus providing further support to the concept of the central node.
However, for the purpose of fitting the observed vertex degree data, 
such nodes were considered outliers. To give an overall measure of 
goodness of the fits, we calculated the $R^2$ {\it coefficient of determination}, 
which can be interpreted as the fraction of the total variation that is 
explained by the least-squares regression line. We obtained, on average, values of $R^2 \approx 0.86$ 
for the entire dataset with outliers included, and $R^2 \approx 0.93$ 
with outliers excluded. Further, the fits for the normal market 
period were better than those obtained for the crash period as 
characterized by the average values of $R^2 \approx 0.89$ and 
$R^2 \approx 0.93$, respectively, with outliers excluded.
In addition to the market period based dependence, 
the exponent $\alpha$ was also found to depend on the window width. 
We examined a range of values for the window width $T$ between 2 
and 8 years and found, without excluding the outliers, the fitted 
exponent to depend linearly on $T$. 

In conclusion, we have found the scaling exponent to depend on 
the market period, i.e. crash vs normal market circumstances and 
on the window width. These results also raise the question of 
whether it is reasonable to assume that different markets share the 
scaling exponent. In case they do not, one should be careful 
when pooling stocks together from different markets for the 
purpose of vertex degree analysis.

\section{Asset tree evolution}

In order to investigate the robustness of asset tree topology, 
we define the \emph{single-step survival ratio} of tree 
edges as the fraction of edges found common in two consecutive trees at 
times $t$ and $t-1$ as

\begin{equation}
\sigma (t)=\frac{1}{N-1}|E(t)\cap E(t-1)|.
\end{equation}

\noindent In this $E(t)$ refers to the set of edges of the tree 
at time $t$, $\cap $ is the intersection operator and $|...|$ gives 
the number of elements in the set. Under normal circumstances, 
the tree for two consecutive time steps should look very similar, 
at least for small values of window step length parameter $\delta T$. 
With this measure it is expected that while some of the differences 
can reflect real changes in the asset taxonomy, others may simply 
be due to noise. On letting $\delta T\rightarrow 0$, we find that 
$\sigma (t)\rightarrow 1$, indicating that the trees \emph{are} 
stable in this limit \cite{jpo}.

\begin{figure}
\epsfig{file=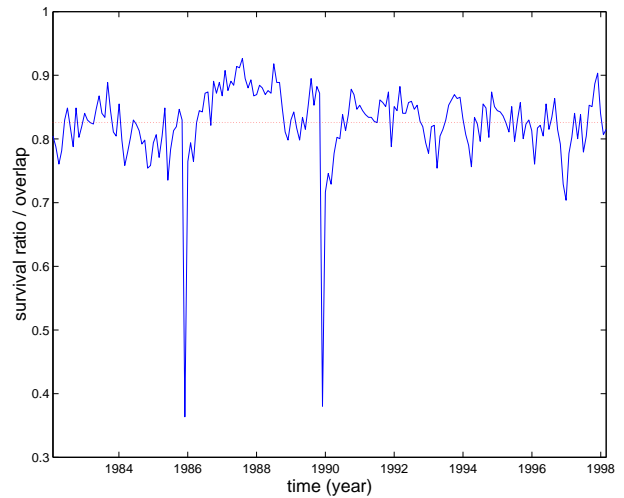,width=3.2in }
\caption{
Single-step survival ratio $\sigma (t)$ as a function of time.
}
\label{sss}
\end{figure}

A sample plot of single-step survival ratio for $T=1000$ and 
$\delta T \approx 20.8$ is shown in Figure \ref{sss}. The 
following observations are made. 
(i) A large majority of connections survives from one time 
window to the next. (ii) The two prominent dips indicate a 
strong tree reconfiguration taking place, and they are window 
width $T$ apart, positioned symmetrically around Black Monday, 
and thus imply topological reorganization of the tree during the market
crash\cite{bali}. (iii) Single-step survival ratio $\sigma (t)$ 
increases as the window width $T$ increases while $\delta T$ is 
kept constant. Thus an increase in window width renders the trees 
more stable with respect to single-step survival of connections. 
We also find that the rate of change of the survival ratio 
decreases as the window width increases and, in the limit, 
as the window width is increased towards infinity 
$T\to \infty ,\quad \sigma (t)\to 1$ for all $t$. 
The survival ratio seems to decrease very rapidly once the 
window width is reduced below roughly one year. As the window 
width is decreased further towards zero, in the limit 
$\textrm{as }T\to 0,\quad \sigma (t)\to 0$ for all $t$. 
(iv) Variance of fluctuations around the mean is constant over 
time, except for the extreme events and the interim period, 
and it gets less as the window width increases.

\begin{figure}
\epsfig{file=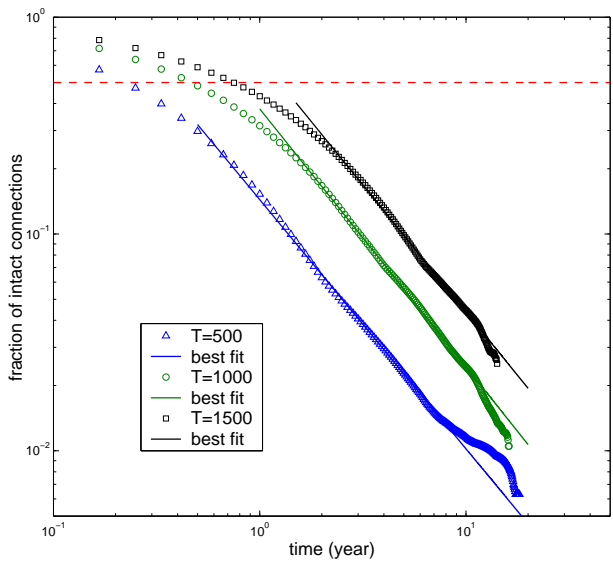,width=3.2in }
\caption{
Multi-step survival ratio $\sigma (t,k)$ as a function of time 
for different parametric values of $T$.
}
\label{mss}
\end{figure}

In order to study the long term evolution of the trees, 
we introduce \emph{the multi-step survival ratio} at time $t$ as 
\begin{equation}
\sigma (t,k)=\frac{1}{N-1}|E(t)\cap E(t-1)...E(t-k+1)\cap E(t-k)|,
\end{equation}
where only those connections that have persisted for the whole time period 
without any interruptions are taken into account. According 
to this formula, when a bond between two companies breaks even 
once in $k$ steps and then reappears, it is not counted as a 
survived connection. It is found that many connections in the 
asset trees evaporate quite rapidly in the early time horizon. 
However, this rate decreases significantly with time, and even 
after several years there are some connections that are left 
intact. This indicates that some companies remain closely 
bonded for times longer that a decade. The behavior of 
the multi-step survival ratio for three different values 
of window width (2,4 and 6 years) is shown in Figure \ref{mss}, 
together with the associated fits.

In this figure the horizontal axis can be divided into two regions. 
Within the first region, decaying of connections is roughly 
exponential, and takes place at different rates for different 
values of the window width. Later, within the second region, 
when most connections have decayed and only some 20\%-30\% remain 
(for the shown values of $T$), there is a cross-over to power 
law behavior. The exponents obtained for the window widths of 
$T=500$, $T=1000$ and $T=1500$ are -1.15, -1.19 and -1.17, respectively. 
Thus, interestingly, the power law decay in the second region 
seems independent of the window width. 

\begin{figure}
\epsfig{file=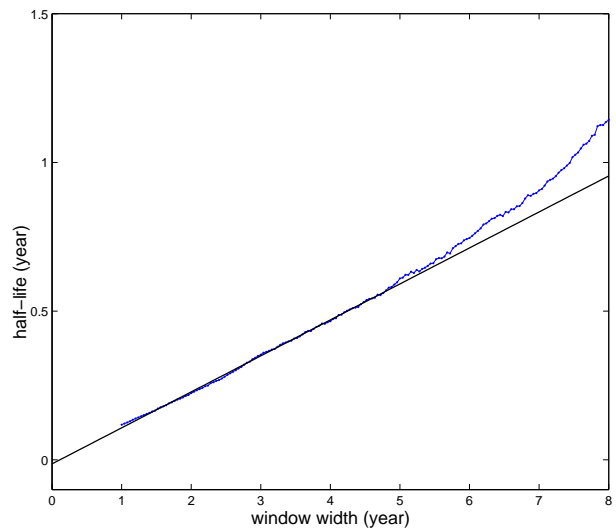,width=3.2in }
\caption{
Plot of half-life $t_{1/2}$ as a function of window width $T$.
}
\label{hl}
\end{figure}

We can also define a characteristic time, the so called half-life 
of the survival ratio $t_{1/2}$, or \emph{tree half-life} for 
short, as the time interval in which half the number of initial 
connections have decayed, i.e., $\sigma (t,t_{1/2})=0.5$. The 
behavior of $t_{1/2}$ as a function of the window width is 
depicted in Figure \ref{hl} and it is seen to follow a clean 
linear dependence on for values of $T$ being between 1 and 5 
years, after which it begins to grow faster than a linear function. 
For the linear region, the tree half-life exhibits 
$t_{1/2} \approx 0.12T$ dependence.

This can also be 
seen in Figure \ref{mss}, where the dashed horizontal line 
indicates the level at which half of the connections have 
decayed. For the studied values of the window width, tree 
half-life occurs within the first region of the multi-step 
survival plot, where decaying was found to depend on the window 
width. Consequently, the dependence of half-life on window 
width $T$ does not contradict the window width independent power 
law decaying of connections, as the two occur in different 
regions. In general, the number of stocks $N$, as well as 
the their type, is likely to affect the half-lives. Earlier, 
for a set of $N=116$ S\&P 500 stocks, half-life was found to 
depend on the window width as $t_{1/2} \approx 0.20T$ \cite{jpo}. 
A smaller tree, consisting primarily of important industry 
giants, would be expected to decay more slowly than the larger 
set of NYSE-traded stocks studied in this paper.

\section{Portfolio analysis}

Next, we apply the above discussed concepts and measures to 
the portfolio optimization problem, a basic problem of financial analysis. 
This is done in the hope that the asset tree could 
serve as another type of quantitative approach to and/or visualization 
aid of the highly inter-connected market, thus acting as a tool 
supporting the decision making process.
We consider a \emph{general Markowitz portfolio} $\mathbf{P}(t)$ with 
the asset weights $w_{1},\, w_{2},\ldots ,\, w_{N}$. In the classic
Markowitz portfolio optimization scheme, financial assets are 
characterized by their average risk and return, where the risk 
associated with an asset is measured by the standard deviation of returns. 
The Markowitz optimization is usually carried out by using historical data.
The aim is to optimize the asset weights so that the overall portfolio 
risk is minimized for a given portfolio return $r_{\mathbf{P}}$ \cite{soft}. 
In the dynamic asset tree framework, however, the task is to 
determine how the assets are located with respect to the 
central vertex. 

Let $r_{m}$ and $r_{M}$ denote the returns of the minimum and 
maximum return portfolios, respectively. The expected portfolio 
return varies between these two extremes, and can be expressed 
as $r_{\mathbf{P},\theta}=(1-\theta) r_{m} + \theta r_{M}$, 
where $\theta$ is a fraction between 0 and 1. Hence, 
when $\theta = 0$, we have the minimum risk portfolio, 
and when $\theta = 1$, we have the maximum return (maximum risk) 
portfolio. The higher the value of $\theta$, the higher the 
expected portfolio return $r_{\mathbf{P},\theta}$ and, 
consequently, the higher the risk the investor is willing to 
absorb. We define a single measure, the 
\emph{weighted portfolio layer} as

\begin{equation}
l_{\mathbf{P}}(t,\theta)=\sum _{i\in \mathbf{P}(t,\theta)}w_{i}\mathop {\mathrm{lev}}(v_{i}^{t}),\end{equation}

\noindent where $\sum_{i=1}^{N}w_i=1$ and further, as a starting point, 
the constraint $w_{i}\geq 0$ for all $i$, which is equivalent to 
assuming that there is no short-selling. The purpose of this constraint 
is to prevent negative values for $l_{\mathbf{P}}(t)$, which would 
not have a meaningful interpretation in our framework of trees with 
central vertex. This restriction will shortly be discuss further.

\begin{figure}
\epsfig{file=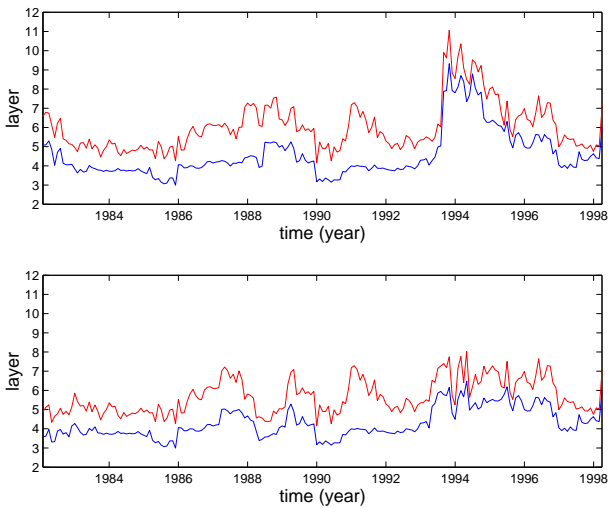,width=3.2in }
\caption{
Plot of the weighted minimum risk portfolio layer 
$l_{\mathbf{P}}(t,\theta = 0)$ with no short-selling and 
mean occupation layer $l(t,v_{c})$ against time. 
Top: static central vertex, bottom: dynamic central vertex 
according to the vertex degree criterion.
}
\label{portnss}
\end{figure}

Figure \ref{portnss} shows the behavior of the mean occupation 
layer $l(t)$ and the weighted minimum risk portfolio layer 
$l_{\mathbf{P}}(t,\theta = 0)$. 
We find that the portfolio layer is higher than the mean layer at 
all times. The difference between the layers depends on the window 
width, here set at $T=1000$, and the type of central vertex 
used. The upper plot in Figure \ref{portnss} is produced using 
the static central vertex (GE), and the difference in layers is found 
to be 1.47. The lower one is produced by using a dynamic central 
vertex, selected with the vertex degree criterion, in which case 
the difference of 1.39 is found. 

\begin{figure}
\epsfig{file=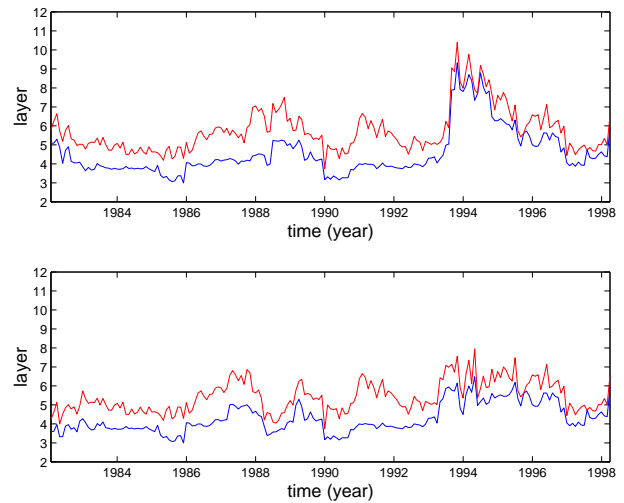,width=3.2in }
\caption{
Plot of the weighted minimum risk portfolio layer 
$l_{\mathbf{P}}(t,\theta = 0)$ with short-selling allowed 
and mean occupation layer $l(t,v_{c})$ against time. Top: static 
central vertex, bottom: dynamic central vertex according to 
the vertex degree criterion.
}
\label{portssa}
\end{figure}

Above we assumed the no short-selling condition. However, it turns 
out that, in practice, the weighted portfolio layer never assumes 
negative values and the short-selling condition, in fact, is not 
necessary. Figure \ref{portssa} repeats the earlier plot, this 
time allowing for short-selling. The weighted portfolio layer is 
now 99.5\% of the time higher than the mean occupation layer and, 
with the same central vertex configuration as before, the difference 
between the two is 1.18 and 1.14 in the upper and lower plots, 
respectively. Thus we conclude that only minor differences are 
observed in the previous plots between banning and allowing 
short-selling, although the difference between weighted portfolio 
layer and mean occupation layer is somewhat larger in the first 
case. Further, the difference in layers is also slightly larger 
for static than dynamic central vertex, although not by much.

As the stocks of the minimum risk portfolio are found on the 
outskirts of the tree, we expect larger trees (higher $L$) to 
have greater \emph{diversification potential}, i.e., the scope 
of the stock market to eliminate specific risk of the minimum 
risk portfolio. In order to look at this, we calculated the 
mean-variance frontiers for the ensemble of 477 stocks using 
$T=1000$ as the window width. In Figure \ref{clr}, we plot 
the level of portfolio risk as a function of time, and find 
a similarity between the risk curve and the curves of the mean 
correlation coefficient $\bar{\rho }$ and normalized tree 
length $L$. Earlier, when the smaller dataset of 116 stocks - 
consisting primarily important industry giants - was used, 
we found Pearson's linear correlation between the 
risk and the mean correlation coefficient $\bar{\rho}(t)$ to be $0.82$, 
while that between the risk and the normalized tree 
length $L(t)$ was $-0.90$. Therefore, for that dataset, 
the normalized tree length was able to explain the 
diversification potential of the market better than the mean 
correlation coefficient. For the current set of 477 stocks, which  
includes also less influential companies, the Pearson's linear 
and Spearman's rank-order correlation coefficients between 
the risk and the mean correlation coefficient are 0.86 and 
0.77, and those between the risk and the normalized tree 
length are -0.78 and -0.65, respectively. It should be noted 
again that the minimum spanning tree with only $N-1$ elements 
represents a pruned version of the entire system of $N(N-1)/2$ 
elements. Further, as $N$ increases, the proportion of elements in 
the tree to the elements in the correlation 
matrix gets less and, consequently, the tree is based on a smaller 
fraction of the available information. Therefore, although 
our earlier finding is not reproduced here to the same extent, 
the result does indicate the strength of pruning the applied 
methodology is able to provide.

\begin{figure}
\epsfig{file=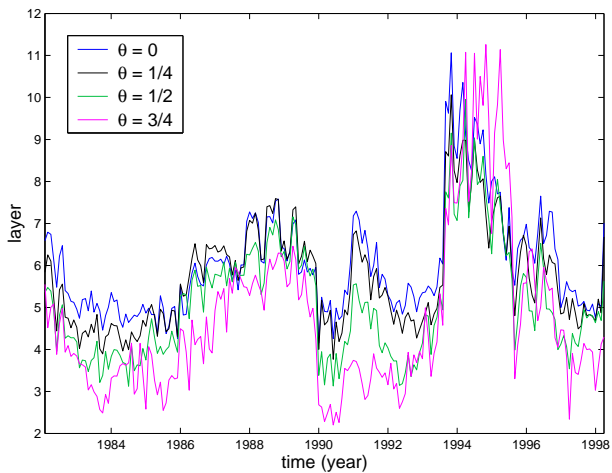,width=3.2in }
\caption{
Plots of the weighted minimum risk portfolio layer
$l_{\mathbf{P}}(t,\theta)$ for different values of $\theta$.
}
\label{port_layers}
\end{figure}

So far, we have only examined the location of stocks in the 
minimum risk portfolio, for which $\theta = 0$. As we increase 
$\theta$ towards unity, portfolio risk as a function of time soon 
starts behaving very differently from the mean correlation 
coefficient and normalized tree length as shown in Fig. \ref{port_layers}. 
Consequently, it is no longer useful in describing diversification 
potential of the market. However, another interesting result 
emerges: The average weighted portfolio layer 
$l_{\mathbf{P}}(t,\theta)$ decreases for increasing values of 
$\theta$. This means that out of all the possible Markowitz 
portfolios, the minimum risk portfolio stocks are located 
furthest away from the central vertex, and as we move towards 
portfolios with higher expected return, the stocks included in 
these portfolios are located closer to the central vertex. 
When static central node is used, the average values of the 
weighted portfolio layer $l_{\mathbf{P}}(t,\theta)$ for 
$\theta = 0, 1/2, 1/2, 3/4$ are 6.03, 5.70, 5.11 and 4.72, 
respectively. Similarly, for a dynamic central node, we obtain 
the values of 5.68, 5.34, 4.78 and 4.37. We have not included the 
weighted portfolio layer for $\theta=1$, as it is not very 
informative. This is due to the fact that the maximum return 
portfolio comprises only one asset (the maximum return asset 
in the current time window) and, therefore, 
$l_{\mathbf{P}}(t,\theta=1)$ fluctuates wildly as the maximum 
return asset changes over time.

We believe these results to have potential for practical application. 
Due to the clustering properties of the MST, as well as the overlap 
of tree clusters with business sectors as defined by a third party 
institution, it seems plausible that companies of 
the same cluster face similar risks, imposed by the external 
economic environment. These dynamic risks influence the stock 
prices of the companies, in coarse terms, leading to their 
clustering in the MST. In addition, the radial location of stocks depends on 
the chosen portfolio risk level, characterized by the value of $\theta$. Stocks included 
in low risk portfolios are consistently located further away from the central node than those 
included in high risk portfolios. Consequently, the radial distance of a node, i.e. its occupation layer, is meaningful. 
Thus, it can be conjectured that the location of a company \emph{within} the 
cluster reflects its position with regard to internal, or cluster 
specific, risk. Characterization of stocks by their branch, 
as well as their location within the branch, enables us to 
identify the degree of interchangeability of different stocks 
in the portfolio.  
For example, in most cases we could pick two stocks from different asset tree clusters, but from nearby layers, and interchange 
them in the portfolio without considerably altering the characteristics of the portfolio. 
Therefore, dynamic asset trees provide an 
intuition-friendly approach to and facilitate 
\emph{incorporation of subjective judgment} in the portfolio 
optimization problem.

\section{Summary and conclusion}

In summary, we have studied the distribution of correlation 
coefficients and found that the mean and the variance of the 
distribution are positively correlated, as well as the skewness 
and the kurtosis. We have also studied the dynamics of asset 
trees and applied it to portfolio analysis. We have shown that 
the tree evolves over time and have found that the normalized 
tree length decreases and remains low during a crash, thus 
implying the shrinking of the asset tree particularly strongly 
during a stock market crisis. We have also found that the mean 
occupation layer fluctuates as a function of time, and experiences 
a downfall at the time of market crisis due to topological 
changes in the asset tree. Further, our studies of the scale 
free structure of the MST show that this graph is not only 
hierarchical in the sense of a tree but there are special, highly 
connected nodes and the hierarchical structure is built up from 
these. As for the portfolio analysis, it was found that the stocks 
included in the minimum risk portfolio tend to lie on the 
outskirts of the asset tree: on average the weighted portfolio 
layer can be almost one and a half levels higher, or further away 
from the central vertex, than the mean occupation layer for window 
width of four years. Correlation between the risk and the normalized 
tree length was found to be strong, though not as strong as the 
correlation between the risk and the mean correlation 
coefficient. Thus we conclude that the diversification potential 
of the market is very closely related to the behavior of the 
normalized tree length. Finally, the asset tree can be viewed 
as a highly graphical tool, and even though it is strongly 
pruned, it still retains all the essential information of 
the market and can be used to add subjective judgment to 
the portfolio optimization problem.

\begin{acknowledgments}
J.-P. O. is grateful to European Science Foundation for REACTOR 
grant to visit Hungary, the Budapest University of Technology 
and Economics for the warm hospitality and Laszlo Kullmann for 
stimulating discussions. Further, the role of Harri Toivonen 
at the Department of Accounting, Helsinki School of Economics, 
is acknowledged for carrying out CRSP database extractions. 
J.-P. O. is also grateful to the Graduate School in Computational 
Methods of Information Technology (ComMIT), Finland.
The authors are also grateful to R. N. Mantegna for very useful 
discussions and suggestions. This research was partially 
supported by the Academy of Finland, Research Center for 
Computational Science and Engineering, project no. 44897 
(Finnish Center of Excellence Programme 2000-2005) and OTKA (T029985). 
\end{acknowledgments}

\appendix
\section{}

\noindent The five sample clusters that were identified in the 
asset tree of Figure \ref{samplegraph} for $t=t^*$, corresponding 
to 1.1.1998, are examined here in closer detail. It is emphasized 
that for purposes of visualization, the tree was constructed from 
a smaller dataset of 116 S\&P 500 stocks. It is also important 
to bear in mind that the words business sector and industry are 
classifications assigned by a third party institution, in 
this case Forbes \cite{forbes}. In contrast, the word cluster 
is used to mean a branch or part of a branch in the tree, where 
most nodes are members of a single business sector.

\noindent\textbf{\emph{Energy cluster}}: 
$L_{\text{Energy}} (t^*) \approx 0.92$. In the dataset there 
are eleven companies operating in the Energy sector, represented 
by red asterisks in Figure \ref{samplegraph}. They form a 
complete Energy cluster, which extends diagonally from the 
center to the bottom left corner of the tree. The industry 
classifications are mainly Oil \& Gas Operations. Only two 
companies, Halliburton (HAL) and Schlumberger (SLB), are 
classified as Oil Well Services \& Equipment. 

\noindent\textbf{\textit{Health-care cluster}}: 
$L_{\text{Health-care}} (t^*) \approx 0.98$. 
The incomplete Health-care cluster extends from the center 
towards the upper left corner of the tree. All seven Health-care 
sector companies, Pfizer (PFE), Eli Lilly (LLY), Merck \& Co. (MRK), 
Johnson \& Johnson (JNJ), Bristol-Myers Squibb (BMY), 
American Home Products (AHP) and Pharmacia (PHA), are 
classified in the Major Drugs industry. As the remaining four 
health care companies operate in different industries, 
this cluster is complete industry wise. 

\noindent\textbf{\emph{Utilities cluster}}: 
$L_{\text{Utilities}} (t^*) \approx 1.01$.
A total of thirteen companies belong to the Utilities business 
sector, represented by the blue asterisks. Twelve of them can 
be found in the incomplete Utilities cluster, which extends 
diagonally from the center to the top right corner of the tree. 
Williams Companies (WMB) is the only company that is not part 
of it, but is located in a sibling branch instead. WMB along 
with Peoples Energy (PGL) are assigned to the Natural Gas 
Utilities industry, where as all other Utilities sector companies 
are assigned to Electric Utilities industry. This can explain 
why WMB is not part of the main branch in the tree.

\noindent\textbf{\textit{Basic Materials cluster}}: 
$L_{\text{Basic materials}} (t^*) \approx 1.03$.
There are thirteen companies in the Basic Materials sector, 
eleven of which are members of the branch on the right hand 
side of the tree. In the incomplete Basic Materials cluster,
we can identify a smaller sub-branch comprising Alcoa (AA), Phelps
Dodge (PD), Homestake Mining (HM) and Inco (N). AA, PD and N are in
the Metal Mining industry and HM in the Gold \& Silver industry. These
are the only four companies within the Basic Materials sector that
provide mining raw materials. Another interesting sub-branch is that
of Georgia-Pacific Group (GP), Weyerhaeuser (WY), Louisiana-Pacific
(LPX) and Boise Cascade (BCC). These companies function in the 
strongly related industries of Paper \& Paper Products and 
Forestry \& Wood Products. We can identify one more sub-branch, 
namely the connected pair of DuPont de Nemours (DD) and Dow 
Chemical Company (DOW), located at the beginning of the
main Basic Materials branch. Both companies are in the Chemicals 
Plastics\& Rubber industry. In the Basic Materials cluster, 
the are three companies included that have a different business 
sector classification from Basic Materials. Two of them, 
Caterpillar (CAT) and Deere \& Company (DE), belong to the 
Capital Goods business sector and Construction \& Agricultural 
Machinery industry. Their position in the branch can be 
substantiated by their reliance on this cluster for raw materials.
The third exception in the Basic Materials sector is International 
Paper (IP), which is located in front of the GP-WY-(LPX,BCC) 
sub-branch. IP belongs to the the Consumer/Non-Cyclical sector 
and within that to the Office Supplies industry. Again, it seems 
natural that a paper company should be located together with 
companies that provide its basic materials.

\noindent\textbf{\textit{Technology cluster}}: 
$L_{\text{Technology}} (t^*) \approx 1.07$.
An example of a clearly incomplete cluster is a group of five 
Technology business sector companies extending diagonally
from the center towards the bottom right corner. These five technology
giants, IBM (IBM), Texas Instruments (TXN), Hewlett-Packard (HWP),
Computer Sciences Corp. (CSC) and Motorola (MOT) form the 
Technology cluster. There are eight other technology companies 
(by business sector) in the set of companies studied, but they 
are mainly distributed around General Electric. The five companies 
of the Technology cluster are grouped together most probably 
because of their involvement with semiconductor industry. Their 
industries are either Semiconductors or Computer Hardware and Computer
Services. Motorola as one of the most important mobile phone 
manufacturers is classified industry-wise as  Communications 
Equipment, a field where similar competencies are required 
as in the previous two.


\begin{thebibliography}{10}
\bibitem{soft}Several software packages based on standard 
procedures are available.
We used \mbox{\textsc{Matlab}} with Financial Toolbox.
\end{thebibliography}
\end{document}